\documentclass[12pt]{article} 
\usepackage{indentfirst}
\usepackage{mathrsfs}
\usepackage{amsmath}
\usepackage{booktabs}
\usepackage{makecell}
\usepackage{graphicx}
\usepackage{longtable}
\usepackage{subfigure}
\usepackage{apacite}
\usepackage[font=footnotesize]{caption}
\usepackage{float}
\usepackage[a4paper,left=2.54cm,right=2.54cm,top=2.54cm,bottom=2.54cm]{geometry}
\usepackage{authblk}
\usepackage[T1]{fontenc}
\usepackage[utf8]{inputenc}
\usepackage{abstract}

\newcommand\keywords[1]{\textbf{Keywords}: #1}

\title{COMEX Copper Futures Volatility Forecasting: Econometric Models and Deep Learning}
\author[1]{Zian Wang}
\author[2]{Xinyi Lu}
\affil[1]{Financial Technology Thrust, The Hong Kong University of Science and Technology, Guangzhou, China}
\affil[2]{Depatment of Statistics, The Chinese University of Hong Kong, Hong Kong, China}

\begin{document}
\maketitle
\begin{abstract}
    This paper investigates the forecasting performance of COMEX copper futures realized volatility across various high-frequency intervals using both econometric volatility models and deep learning recurrent neural network models. The econometric models considered are GARCH and HAR, while the deep learning models include RNN (Recurrent Neural Network), LSTM (Long Short-Term Memory), and GRU (Gated Recurrent Unit). In forecasting daily realized volatility for COMEX copper futures with a rolling window approach, the econometric models, particularly HAR, outperform recurrent neural networks overall, with HAR achieving the lowest QLIKE loss function value. However, when the data is replaced with hourly high-frequency realized volatility, the deep learning models outperform the GARCH model, and HAR attains a comparable QLIKE loss function value. Despite the black-box nature of machine learning models, the deep learning models demonstrate superior forecasting performance, surpassing the fixed QLIKE value of HAR in the experiment. Moreover, as the forecast horizon extends for daily realized volatility, deep learning models gradually close the performance gap with the GARCH model in certain loss function metrics. Nonetheless, HAR remains the most effective model overall for daily realized volatility forecasting in copper futures.
\end{abstract}


\section{Introduction}

As a metal of economic importance, the price of copper is often seen by the industry as a macroeconomic barometer. When copper prices rise, the world economy is in a state of expansion, and vice versa. China's rapidly booming economy has been a major reason for the rise in copper prices for quite some time in the past. And the vigorously developing real estate, the backbone of China's economic boom over the past few decades, has accounted for a significant share of the demand side of the international copper metal trading market. After the impact of the epidemic on the world economy, China's new energy vehicle market is seen as a new demand export for copper products. The suitable climate, dense urban distribution, and transportation characteristics have made investors unanimously bullish on the Chinese tram market. Then future trading of copper futures for hedging in the international market will undoubtedly become more frequent. Therefore, an in-depth study of its high-frequency data volatility forecasts is beneficial to inform the appropriate choices for future investors. In this paper, we analyze the characteristics of COMEX copper futures volatility forecasting from different directions based on the comparison of econometric volatility models and deep learning models.

Although the traditional econometric models are the current commonly used methods to predict volatility, many novel deep learning models show strong promise for better predicting stock behavior (see, e.g., Bucci, 2020; Xiong et al., 2015). Advanced econometric models, such as HAR (heterogeneous autoregression), realized GARCH, HEAVY (high-frequency-based volatility), and Markov-switching GARCH (Conrad and Kleen, 2019), have all been widely applied to volatility forecasting. Different econometric models exhibit specialized treatment of both long-term trends and short-term structural changes.

Compared to econometric models proposed to solve different problems, deep learning models represented by RNN (recurrent neural networks) are more flexible and tunable, which is one of the reasons why they are widely used in high frequency trading. Based on the characteristics of econometric models and deep learning, exploring their prediction effects on different frequency data, and analyzing the corresponding advantages and disadvantages are beneficial to help us make appropriate model choices when facing data-oriented cases.

For the final comparison of volatility forecasting, Patton (2011) finds that both the MSE and the QLIKE loss function are robust when used to compare volatility forecasting models. Rahimikia and Poon (2020) suggested using evaluation measures such as MSE, QLIKE, MDA, and RC values to examine the performance of machine learning models in realized volatility forecasting. This paper uses MSE and QLIKE as criteria for comparing the predictive power of volatility.

\subsection{Copper price is an leading indicator of global economy}

Copper futures are used by miners and dealers to hedge against losses and are in an important position in futures markets all over the world. As an effective conductor of heat and electricity, most copper is used in the electrical sector such as building transmission lines and motors equipment. In the past two decades, the demand of copper has been increasing in developing countries for basic grid construction. As the largest copper importer, China has a huge potential market for electric cars since the production of electric cars requires abundant copper as a raw material. Therefore, a stable and sufficient supply of copper is key to guaranteeing the ongoing shift from gasoline to electric vehicles. Generally, the demand of copper in most developed countries is dominated by manufacturing industry and renewable energy, since it is used to make manufacturing machines and equipment, such as windmills and solar plants. Consequently, copper prices are principally determined by demand of emerging markets such as China and India. 

As copper is associated with many industries, it is often regard as the leading indicator of the world economy. When the global economy is expanding, the copper price tends to rise and vice versa. In recent years, the prices of copper and its derivatives have been fluctuating. While global stock markets have been in recession for the last few years during the pandemic, a soaring rise in copper price took place in 2020. The first reason for the unusual performance is that copper prices are settled in the US dollar, which has fallen sharply during that time. Second, production in major copper exporters such as Chile and Peru have declined due to the COVID-19 pandemic. Thirdly, as having always been the world's largest copper importer for a long time, China did well in the early stages of the pandemic and maintained a sound economic circumstance.

In general, copper has both commodity properties and financial properties, both of which are affected by macroeconomy. Meanwhile, due to the continuous prominence of the financial attributes of the US's commodities, the price volatility of commodity futures market and the price volatility of stock market have increasingly prominent characteristics of synchronization, and the price volatility risk of commodity futures market gradually spreads to the stock market. Therefore, the copper future price is not only affected by the macroeconomic cycle, but also has an increasing correlation with the stock market. 

\subsection{Anticipating volatility based on econometric models}

As a statistical measure of the dispersion of returns for a given security or market index in finance, volatility is often used as a measure of risk. Higher volatility means that the price of an asset can fluctuate dramatically in a short period of time, which can lead to large gains or losses for investors. Volatility models are used to forecast the absolute magnitude of returns and predict quantiles or the entire density of future returns. There are many types of volatility models, including stochastic volatility models which assume that the volatility itself follows a random process. Moreover, realized volatility is a measure of the variation in the price of an asset over a given period. Realized volatility models are used to forecast future volatility using past realized volatility. The strength of realized volatility is that it is based on actual prices and is therefore less susceptible to errors due to assumptions about the underlying distribution of returns. This paper only calculates the realized volatility of COMEX copper futures based on daily or high-frequency intraday return respectively. 

HAR (Heterogeneous Autoregressive) uses the squared past returns to predict future volatility (Corsi, 2009). The realized GARCH incorporates the HAR structure of realized variance into the GARCH equation (Hansen et al., 2012). ARFIMA (Autoregressive Quantile Integrated Moving Average) captures the long-term memory in volatility (Hosking, 1981). The HEAVY (high-frequency-based volatility) model is used to capture heavy tails in financial data (Shephard and Sheppard, 2010). The MS-GARCH (Markov-Switching GARCH) allows the volatility regime to vary over time (Hamilton and Susmel, 1994). This paper chooses HAR and realized GARCH as representatives of econometric models because of stationarity and applicability. Both of them have good performance in predicting realized volatility.

\subsection{Deep Learning models have potential in volatility forecasting}

Forecasting realized volatility is essential for trading signals and position management. Econometric models, such as GARCH and HAR, predict future volatility based on past returns in an intuitive way. Conrad and Kleen (2019) stated that GARCH-MIDAS outperforms several econometric competitor models such as HAR, HEAVY, realized GARCH, and MS-GARCH. However, recurrent neural networks have become a significant competitor, especially LSTM and GRU. The discussion on whether traditional econometric models (such as GARCH) or deep learning models (such as LSTM) have higher prediction accuracy has not stopped. 

Xiong et al. (2016) applied a Long Short-Term Memory neural network to model S$\&$P 500 volatility, and its model has a smaller mean absolute percentage error compared with linear Ridge/Lasso and autoregressive. Bucci (2019) compared the predictive performance of feed-forward and recurrent neural networks (RNN) with traditional econometric approaches, and the RNN outperforms all the traditional econometrics methods. Rahimikia and Poon (2020) showed that the machine learning model represented by LSTM has a stronger prediction ability than all HAR-family models based on machine learning models to predict volatility strengths and weaknesses for 23 Nasdaq stocks between 2007 and 2016. Despite the increasing achievements of machine learning in volatility prediction, it is still too early to say that classical econometric models will be replaced.

MSE and RMSE are used to measure the average squared difference between the predicted and actual values. MAPE is used to measure the percentage difference between the predicted and actual values. MAE is used to measure the average absolute difference between the predicted and actual values. QLIKE is used to measure the difference between the predicted and actual quantiles. The choice of which metric to use depends on the specific problem at hand. MSE and RMSE are commonly used in regression problems, while MAPE and MAE are commonly used in forecasting problems. QLIKE is commonly used in quantile regression problems. This paper analyzes the forecasting results of RNN (Recurrent Neural Network), LSTM (Long Short-Term Memory), and GRU (Gated Recurrent Unit) for realized volatility of copper futures at different frequencies and compare them with the loss functions of the above econometric models over time. According to Patton (2011), this paper collects MSE, RMSE, MAPE, MAE and QLIKE loss function to compare robust loss validation trend of different models.

\section{Methodology}
\subsection{Econometric Models}
\subsubsection{GARCH(1,1)}

Engle and Lee(1999) introduced a GARCH model with a long-run and a short-run component. GARCH(1,1) represents that the volatility is based on the most recent observation of return and the most recent estimate of the variance rate. Furthermore, GARCH(p,q) means p observations of return and q estimates of variance rate.

\begin{equation}
    \sigma_{n}^2 = \gamma V_{L} + \alpha u_{n-1}^2 + \beta \sigma_{n-1}^2 \tag{1} \\
\end{equation}
$$
\omega = \gamma V_{L}
$$
$$
\gamma + \alpha + \beta = 1
$$


\subsubsection{Realized GARCH}

Realized GARCH framework is a joint modelling of return and realized volatility. The general structure of Realized GARCH (p,q) model is given by

\begin{equation}
    RV_t = \sum_{i = 1}^{N_t} r^2_{i,t} \tag{2}
\end{equation}
\begin{equation}
    r_{t} = \sqrt{h_{t}} z_{T} \tag{3}
\end{equation}
\begin{equation}
    h_{t} = \omega + \alpha r_{t-1}^2 + \beta h_{t-1} + \gamma x_{t-1} \tag{4}
\end{equation}
$$
x_{t} = \xi + \phi h_{t} + Error_{t},
$$
where $x_{t}$ is noisy mearsurement of $QV_{t}$, and $QV_{t}$ is $h_{t}+$ volatility shock.

\subsubsection{HAR}

With the widespread availability of high-frequency intraday data, the HAR model proposed by Corsi(2009) has gained popularity due to its simplicity and consistent forecasting performance in applications. The conditional variance of the discretely sampled returns is parameterized as a linear function of lagged squared returns over the same horizon together with the square returns over longer and shorter horizons.

The original HAR model specifies RV as a linear function of daily, weekly and monthly realized variance components, and can be expressed as

\begin{equation}
    RV_{t} = \beta_{0} + \beta_{1} RV_{t-1}^{d} + \beta_{2} RV_{t-1}^{\omega} + \beta_{3} RV_{t-1}^{m} + u_{t}, \tag{5}
\end{equation}
where the $\beta_{j}(j=0,1,2,3)$ are unknown parameters that need to be estimated, $RV_{t}$ is the realized variance of day $t$. The Specification of RV parsimoniously captures the high persistence observed in most realized variance series.

\subsection{Deep Learning}
The model used in this paper aims to predict volitality of COMEX based on different neural networks. The input data is preprocessed through normalization using the MinMaxScaler. LSTM(long short term memory), GRU (gated recurrent unit) and RNN (recurrent neural network) enhance the modeling of sequential data by explicitly considering the dependencies between observations when learning the mapping relationship from input to output, which allows them to capture the sequential dependencies in the input data.

The dataset is divided into training and testing sets, and input-output pairs for training are generated by sliding a window of 12 consecutive days in the training set because the recurrent model expect the input component of training data to have the dimensions of [batch samples, time sequence length, number of features]. The model architecture is constructed using the Keras Sequential API. For the GRU model, the input layer contains 16 units. This is followed by a dense layer with 4 units and a LeakyReLU activation function to introduce non-linearity, and another dense layer with the number of output days as units is added. The model is then compiled using the Adam optimizer with a learning rate of 0.0001, and the mean squared error is selected as the loss function. Training is performed for 50 epochs with a batch size of 64. To prevent overfitting, a dropout layer with a dropout rate of 0.2 is incorporated,. and early stopping based on the model loss is also employed. In the LSTM model, the input layer consists of 8 units, followed by a dense layer with the number of output days as units. The batch size has been adjusted to 16, while the remaining architecture and hyperparameters remain unchanged. The RNN network has the same structure as the LSTM network, but the number of epochs has been set to 30.

It can try to change the sequence length to 50 because the 50-day moving average is a commonly used technical indicatorthat provides insights into the trend and momentum of assets. Traders actually utilize the 50-day moving average to confirm price trends, determine support and resistance levels, and generate trading signals.
\subsubsection{RNN}

RNNs have the same input and output architecture as any other deep neural architecture. However, differences arise in the way information flows from input to output. Unlike Deep neural networks where we have different weight matrices for each Dense network in RNN, the weight across the network remains the same. It calculates state hidden state $H_{i}$ for every input $X_{i}$. By using the following formulas:

\begin{equation}
    h=\sigma(UX+Wh_{-1}+B) \tag{6}
\end{equation}
\begin{equation}
    Y=O(Vh+C) \tag{7}
\end{equation}
\begin{equation}
    Y=f(X,h,W,U,V,B,C). \tag{8}
\end{equation}

Here S is the State matrix which has element $s_{i}$ as the state of the network at timestep $i$. The parameters in the network are $W$, $U$, $V$, $c$, $b$ which are shared across timestep. The formulas for calculating the current state, applying Activation function(tanh), and calculating output are given below respectively.

\begin{equation}
    h_{t} = f(h_{t-1},x_{t}) \tag{9}
\end{equation}
\begin{equation}
    h_{t} = tanh(W_{hh}h_{t-1}+W_{xh}x_{t}) \tag{10}
\end{equation}
\begin{equation}
    y_{t} = W_{hy}h_{t}, \tag{11}
\end{equation}
where $x_{t}$ means input state; $w_{hh}$, $w_{xh}$, and $w_{hy}$ represent weights at recurrent neuron, input neuron and output layer.
\subsubsection{LSTM and GRU}
The basic difference between the architectures of RNNs and LSTMs is that the hidden layer of LSTM is a gated unit or gated cell. It consists of four layers that interact with one another in a way to produce the output of that cell along with the cell state. These two things are then passed onto the next hidden layer. Unlike RNNs which have got the only single neural net layer of tanh, LSTMs comprises of three logistic sigmoid gates and one tanh layer. Gates have been introduced in order to limit the information that is passed through the cell. They determine which part of the information will be needed by the next cell and which part is to be discarded. The output is usually in the range of 0-1 where '0' means 'reject all' and '1' means 'include all'.  

Each LSTM cell has three inputs $h_{t-1}$, $C_{t-1}$,  and $x_{t}$  and two outputs $h_{t}$  and $C_{t}$. For a given time $t$, $h_{t}$ is the hidden state, $C_{t}$  is the cell state or memory, $x_{t}$  is the current data point or input. The first sigmoid layer has two inputs–$h_{t-1}$  and $x_{t}$  where $h_{t-1}$  is the hidden state of the previous cell. It is known as the forget gate as its output selects the amount of information of the previous cell to be included. The output is a number in [0,1] which is multiplied (point-wise) with the previous cell state $C_{t-1}$. 

\begin{figure}[H]
    \centering
    \caption{Deep Learning Models}
    \includegraphics[width=\textwidth]{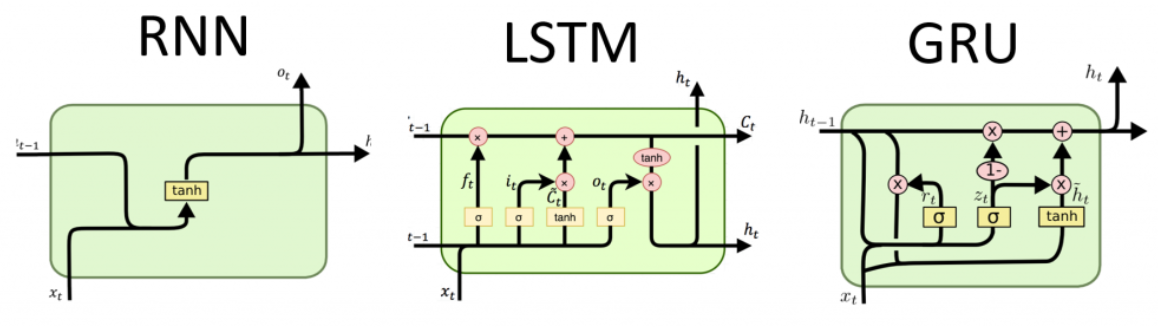}
\end{figure}

In spite of being quite similar to LSTMs, GRUs have never been so popular. GRU stands for Gated Recurrent Units. As the name suggests, these recurrent units, proposed by Cho, are also provided with a gated mechanism to effectively and adaptively capture dependencies of different time scales. They have an update gate and a reset gate. The former is responsible for selecting what piece of knowledge is to be carried forward, whereas the latter lies in between two successive recurrent units and decides how much information needs to be forgotten. 

\begin{equation}
    \text{Activation at time t: } h_{t}^{j} = (1-z_{t}^{j})h_{t-1}^{j}+z_{t}^{j}*\tilde{h}_{t}^{j}  \tag{12}
\end{equation}
\begin{equation}
    \text{Update gate: } z_{t}^{j} = \sigma(W_{z}x_{t}+U_{z}h_{t-1})^{j}  \tag{13}
\end{equation}
\begin{equation}
    \text{Candidate activation: } \tilde{h}_{t}^{j} = tanh(Wx_{t}+U(r_{t}\otimes h_{t-1}))^{j} \tag{14}
\end{equation}
\begin{equation}
    \text{Reset gate: } r_{t}^{j} = \sigma(W_{r}x_{t}+U_{r}h_{t-1})^{j} \tag{15}
\end{equation}

Another striking aspect of GRUs is that they do not store cell state in any way, hence, they are unable to regulate the amount of memory content to which the next unit is exposed. Instead, LSTMs regulate the amount of new information being included in the cell. On the other hand, the GRU controls the information flow from the previous activation when computing the new, candidate activation, but does not independently control the amount of the candidate activation being added (the control is tied via the update gate). 

\subsection{QLIKE Loss Function}

The Quadratic Likelihood Error (QLIKE) loss function is widely used in the evaluation of volatility forecasting models, particularly in the field of finance. It is designed to assess the accuracy of volatility or variance predictions in time series data. Compared to other loss functions such as RMSE or MAPE, QLIKE is more robust to scale differences and highly sensitive to extreme values, making it particularly useful for comparing volatility forecasting models.

Given a set of actual volatilities \( \sigma_t^2 \) and predicted volatilities from a model \( \hat{\sigma}_t^2 \), the QLIKE loss function is defined as follows:

\begin{equation}
    \text{QLIKE} = \frac{1}{T} \sum_{t=1}^{T} \left( \log(\hat{\sigma}_t^2) + \frac{\sigma_t^2}{\hat{\sigma}_t^2} - 1 \right) \tag{16},
\end{equation}
where $T$ represents the length of the time series, $\sigma_t^2$ is the actual volatility (or true variance) at time $t$. And $\hat{\sigma}_t^2$ is the predicted volatility from the model at time \( t \).

If the predicted volatility is close to the actual volatility, the QLIKE value will be close to 0. A higher QLIKE value indicates a larger forecasting error, particularly when the model underestimates volatility. By comparing the QLIKE values of different models, researchers can determine which model performs best in volatility forecasting.
\section{Data Specification}
\subsection{Data Collection}
This article collects daily COMEX copper futures prices from the Wind database from 2000/1/4 to 2023/3/2. Daily log returns are calculated from the daily copper prices, and then assuming M is 1, a preliminary daily realized volatility is obtained as a volatility measure (which is actually the square of the daily return). And for high frequency data, this paper collects the price changes from 2023-01-02 18:01 to 2023-04-13 03:27, which in turn gives the hourly RV (which is actually the sum of the squares of the returns per minute).

\begin{table}[H]
    \footnotesize
    \centering
    \caption{Data Collection}
    \label{Table1}
    \begin{tabular}{lllllllll}
                       & \textbf{Start} & \textbf{End}   & \textbf{Obs.} & \textbf{Window Size} &  &  &                      &                      \\ \hline
    \multicolumn{5}{l}{\textbf{Training Set}}                                                     &  &  & \multicolumn{1}{c}{} & \multicolumn{1}{c}{} \\ 
    Daily Price        & 2000/1/4       & 2023/3/2       & 5777            & /                    &  &  &                      &                      \\
    Daily Return       & 2000/1/5       & 2023/3/2       & 5776            & /                    &  &  &                      &                      \\
    Daily RV           & 2000/1/5       & 2023/3/2       & 5776            & /                    &  &  &                      &                      \\
    1 min Price Change & 2023/1/2 18:01 & 2023/4/13 3:27 & 99999           & /                    &  &  &                      &                      \\
    Houlry RV          & 2023/1/2 20:00 & 2023/4/13 3:00 & 1641            & /                    &  &  &                      &                      \\ \hline
    \multicolumn{5}{l}{\textbf{Test Set}}                                                         &  &  &                      &                      \\ 
    Daily RV           & 2016/5/19 0:00 & 2023/3/2 0:00  & 1688            & 4077                 &  &  &                      &                      \\
    Hourly RV          & 2023/3/15 7:00 & 2023/4/13 3:00 & 480             & 1149                 &  &  &                      &                      \\ \hline
    \end{tabular}
\end{table}

According to the hand-out method, the training set occupies seventy percent of the original data and the test set occupies thirty percent of the original data. This article uses the rolling window to predict the out of sample data and compares and analyzes the loss functions of different models for copper futures volatility predicted data with the real data.

\subsection{Statistical Summary}
Table 2 provides basic descriptive statistics about COMEX copper future’s return and RV in terms of observation, minimum value, maximum value, standard deviation, skewness, and kurtosis. Simultaneously, we choose the Jarque-Bera test to confirm normality. The result is that all of them are not normally distributed. Apart from that, no time series is white noise by the Ljung-Box test lags with 20, and they all have an ARCH effect. All of them are stationary according to ADF test.

\begin{table}[H]
    \footnotesize
    \centering
    \caption{Statistical Summary}
    \label{Table2}
    \resizebox{\textwidth}{!}{%
    \begin{tabular}{clllllccccc}
           & \multicolumn{1}{c}{\textbf{Obs.}} & \multicolumn{1}{c}{\textbf{Mean}} & \multicolumn{1}{c}{\textbf{SD}} & \multicolumn{1}{c}{\textbf{Min}} & \multicolumn{1}{c}{\textbf{Max}} & \textbf{Skew.} & \textbf{Kurt.} & \textbf{J-B pvalue} & \textbf{L-B pvalue} & \textbf{ADF} \\ \hline
    \multicolumn{11}{c}{\textbf{Daily}}                                                                                                                                                                                                                                                 \\ \hline
    Return & 5777                              & 0.0003                            & 0.0168                          & -0.1169                          & 0.1177                           & -0.1772        & 4.1250         & 0.0000              & 0.0000              & 0.0000       \\
    RV     & 5777                              & 0.0003                            & 0.0007                          & 0.0000                           & 0.0139                           & 8.3841         & 112.7821       & 0.0000              & 0.0000              & 0.0000       \\
    \multicolumn{11}{c}{\textbf{Hourly}}                                                                                                                                                                                                                                                \\ \hline
    Return & 1641                              & 4.31E-05                          & 0.0032                          & -0.0177                          & 0.0160                           & -0.0932        & 4.3197         & 0.0000              & 0.0000              & 0.0000       \\
    RV     & 1641                              & 9.78E-06                          & 0.0000                          & 0.0000                           & 0.0001                           & 3.5647         & 23.1754        & 0.0000              & 0.0000              & 0.0000       \\ \hline
    \end{tabular}%
    }
\end{table}
\textit{Note: According to the ADF test, the data is stationary after processing. For the J-B test, if the p-value is bigger than 0.05, the time series will be regarded as normally distributed. For the L-B test, the data is not going to be white noise if the p-value is smaller than 0.05.}

\section{Empirical Analysis}

\subsection{Daily Volatility Forecasting}

For the daily realized volatility forecasts, the rolling window size is set to 4077 in this paper. The out-of-sample predicted data obtained from the Rolling window forecasts are from 2016/5/19 to 2023/3/2, with a total of 1688 data. It can be seen from the picture that both recurrent neural network models (RNN, LSTM, and GRU) and econometric models (realized GARCH and HAR) accurately capture the occurrence of extreme volatility. However, none of the current tested models can be closer to the value of the true extreme volatility. Considering that the original volatility estimation method uses the square of the daily return as the realized volatility, there may be a coarser estimation error. As a result, information on intra-day copper futures price volatility is not captured, making it difficult to approach the true value when extreme volatility occurs. Still, in terms of general trends, the above model is a better predictor of long-term structural changes.

\begin{figure}[H]
    \centering
    \caption{Daily Volatility Forecasting}
    \includegraphics[width=\textwidth]{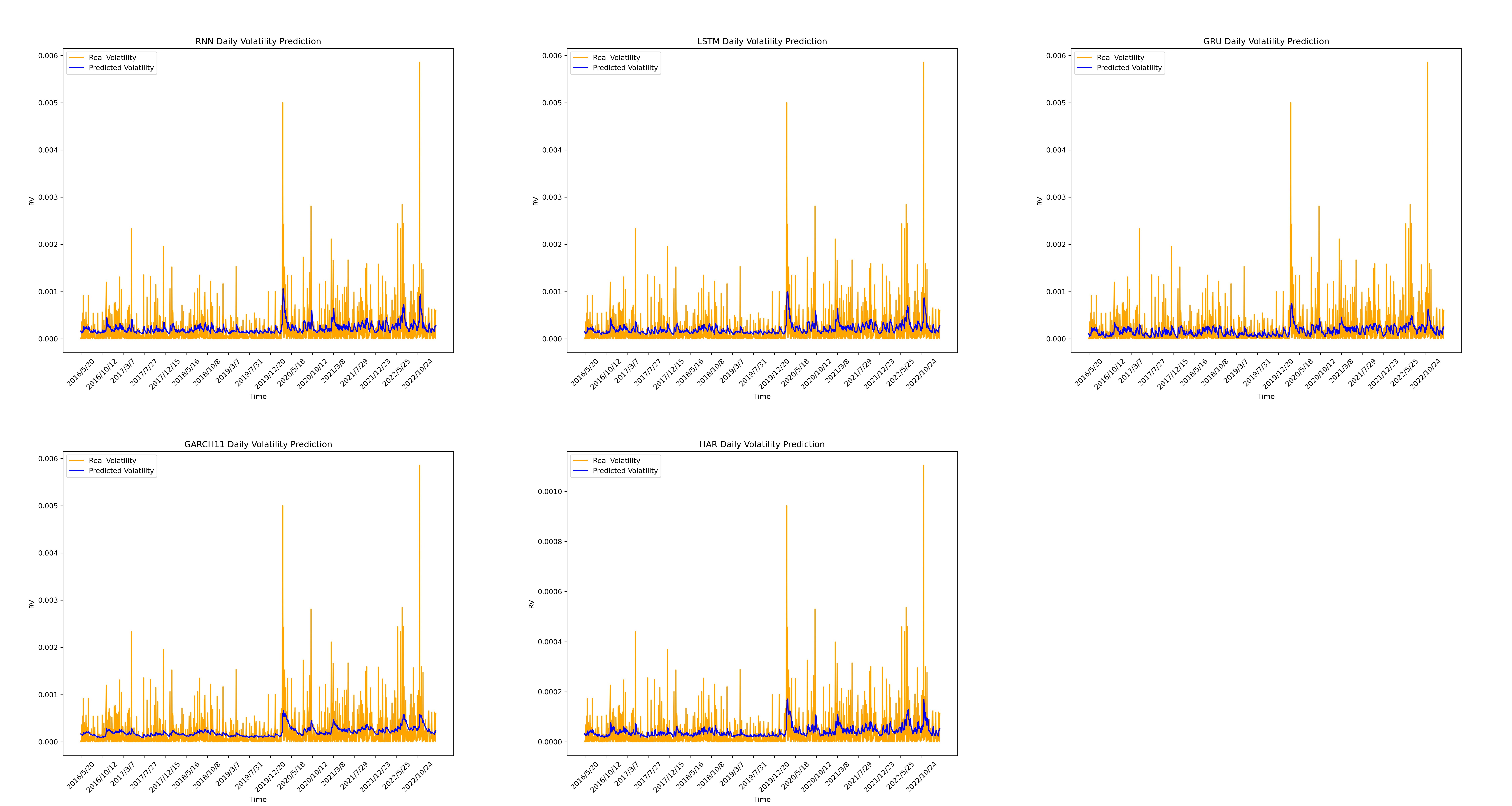}
\end{figure}

As can be seen in the table, the RNN recurrent neural network models are very similar to each other and their performance in the selected data can be considered almost indistinguishable. Since the volatility measure is very small, the value of MSE is not precise enough to be ignored as 0. From the RMSE as well as MAPE, it can be seen that the GARCH model is more prominent in daily volatility forecasting. While MAE, which is widely used in forecasting problems, indicates that HAR performs better in long-term daily frequency volatility forecasting. Combined with the QLIKE loss function, it can be concluded that the econometric models represented by the realized GARCH and HAR perform better than the recurrent neural network models in forecasting the realized volatility of daily frequency copper futures. The econometric volatility models are still much stronger than the deep learning models in predicting long-term trends and capturing structural changes.

\begin{table}[H]
    \footnotesize
    \centering
    \caption{Daily Volatility Forecasting}
    \label{Table3}
    \begin{tabular}{llllll}
                   & \textbf{RNN} & \textbf{LSTM} & \textbf{GRU} & \textbf{rGARCH} & \textbf{HAR} \\ \hline
    \textbf{MSE}   & 0.0000       & 0.0000        & 0.0000       & 0.0000                  & 0.0000       \\
    \textbf{RMSE}  & 0.0004       & 0.0004        & 0.0004       & 0.0003                  & 0.0001       \\
    \textbf{MAPE}  & 1.0194       & 1.0534        & 1.3938       & 0.9250                  & 1.0320       \\
    \textbf{MAE}   & 0.0002       & 0.0002        & 0.0002       & 0.0002                  & 0.0000       \\
    \textbf{QLIKE} & 6.73E-08     & 6.70E-08      & 6.72E-08     & 5.99E-08                & 2.39E-09     \\ \hline
    \end{tabular}
    \end{table}

It is well known that futures trading is often accompanied by dramatic price fluctuations, and how to improve the accuracy of intraday volatility forecasting has become a concern for many traders as well as investors.

\subsection{Hourly Volatility Forecasting}

For hourly realized volatility forecasts, the prize window size is set to 1149 in this paper. Out-of-sample forecasts obtained from rolling window forecasts range from 2023/3/15 7:00 to 2023/4/13 3:00, for a total of 480 data points.

\begin{figure}[H]
    \centering
    \caption{Hourly Volatility Forecasting}
    \includegraphics[width=\textwidth]{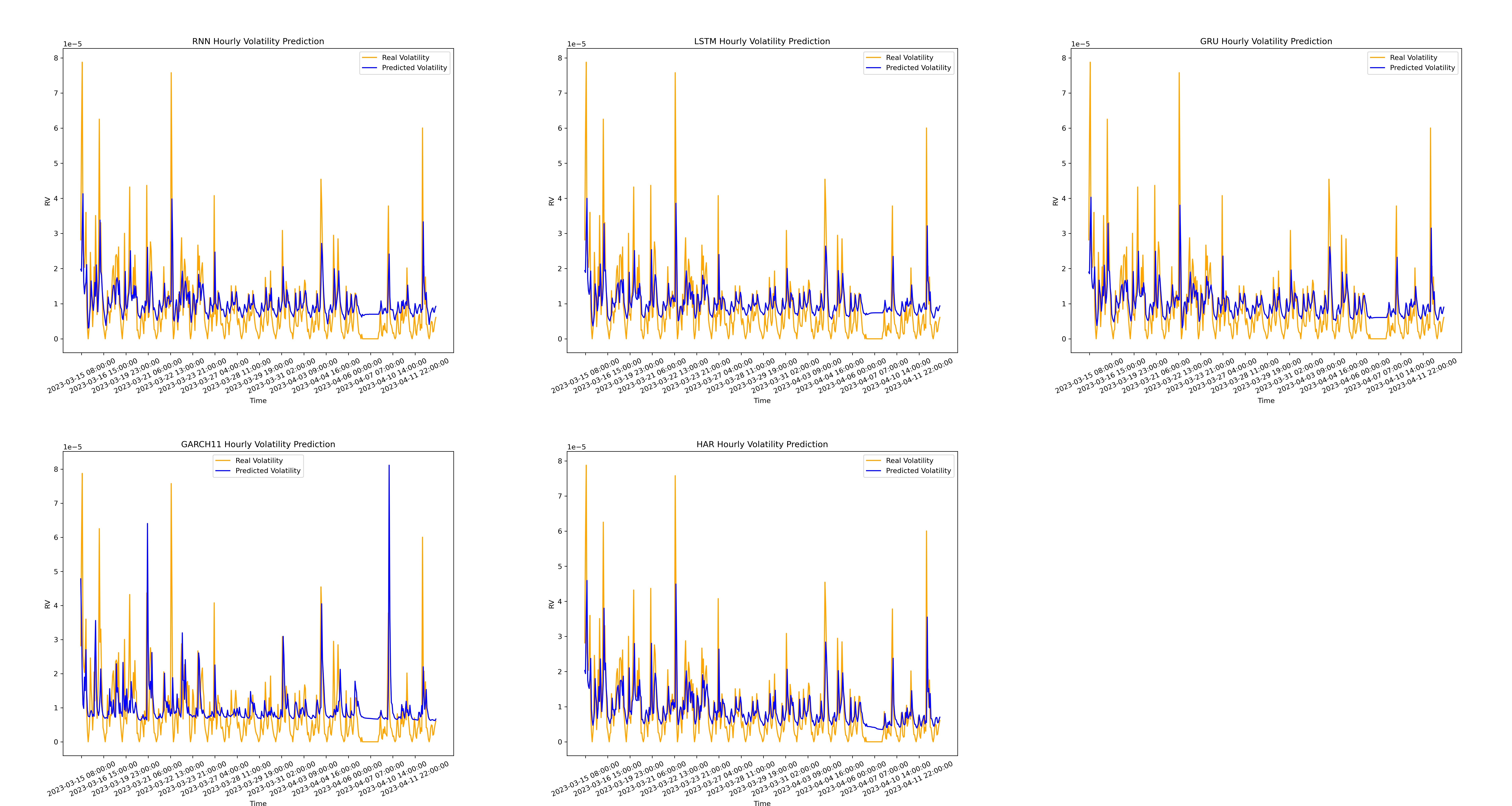}
\end{figure}

Unlike the daily volatility forecasting performance, the recurrent neural network model significantly outperforms the overall GARCH forecasting performance in intra-day forecasting. From the image, the GARCH model is difficult to adjust in time in high frequency data, thus leading to inaccurate results most of the time. In terms of the QLIKE loss function, HAR is the best performing model, but the difference in prediction results with the deep learning model is not significant and can even be ignored. The deep learning model, on the other hand, because of the reason that the results are not exactly the same every time, in some cases it can be believed that the recurrent neural network model has the full potential to outperform the best performing HAR model. This is what has happened in the tests.

\begin{table}[H]
    \footnotesize
    \centering
    \caption{Houlry Volatility Forecasting}
    \begin{tabular}{llllll}
                   & \textbf{RNN} & \textbf{LSTM} & \textbf{GRU} & \textbf{realized GARCH} & \textbf{HAR} \\ \hline
    \textbf{MSE}   & 0.0000       & 0.0000        & 0.0000       & 0.0000                  & 0.0000       \\
    \textbf{RMSE}  & 8E-6         & 9E-6          & 8E-6         & 1E-5                    & 8E-6         \\
    \textbf{MAPE}  & 0.6008       & 0.6129        & 0.6052       & 0.6864                  & 0.6078       \\
    \textbf{MAE}   & 0.000006     & 0.000006      & 0.000006     & 0.000007                & 0.000005     \\
    \textbf{QLIKE} & 3.59E-11     & 3.73E-11      & 3.61E-11     & 5.19E-11                & 3.43E-11     \\ \hline
    \end{tabular}
    \end{table}

Therefore for more frequent volatility forecasting, there is reason to believe that deep learning models are a better choice than traditional econometric volatility models. The econometric volatility model represented by HAR model can be used as benchmark to measure higher frequency volatility forecasts. However, due to the difficulty of collecting data sets, this paper will not consider higher frequency volatility forecasts.

\subsection{Loss Function Trend}

While deep learning models, particularly recurrent neural networks, demonstrate strong performance in forecasting intraday hourly realized volatility, their effectiveness diverges significantly from econometric models when applied to long-term daily frequency volatility forecasts. Econometric models tend to provide more transparent and interpretable solutions, which may be particularly advantageous for forecasting over extended periods. In practice, however, high-frequency traders and fund managers are not only concerned with short-term volatility changes, such as those occurring within a day or the following day, but also with longer-term price fluctuations and volatility trends. While macroeconomic researchers may offer broad, accurate insights, the precise quantification of volatility trends for specific financial products remains challenging. This raises important questions about whether deep learning models continue to underperform in longer-term volatility forecasts when compared to econometric models, particularly in terms of forecast accuracy and actual errors. A thorough comparative analysis of the predictive performance of deep learning models over longer horizons, assuming a 30-day month for forward projections, could offer valuable insights and contribute to a deeper understanding of their applicability in financial volatility forecasting.
\begin{figure}[]
    \centering
    \caption{Loss Function Trend}
    \includegraphics[width=0.8\textwidth]{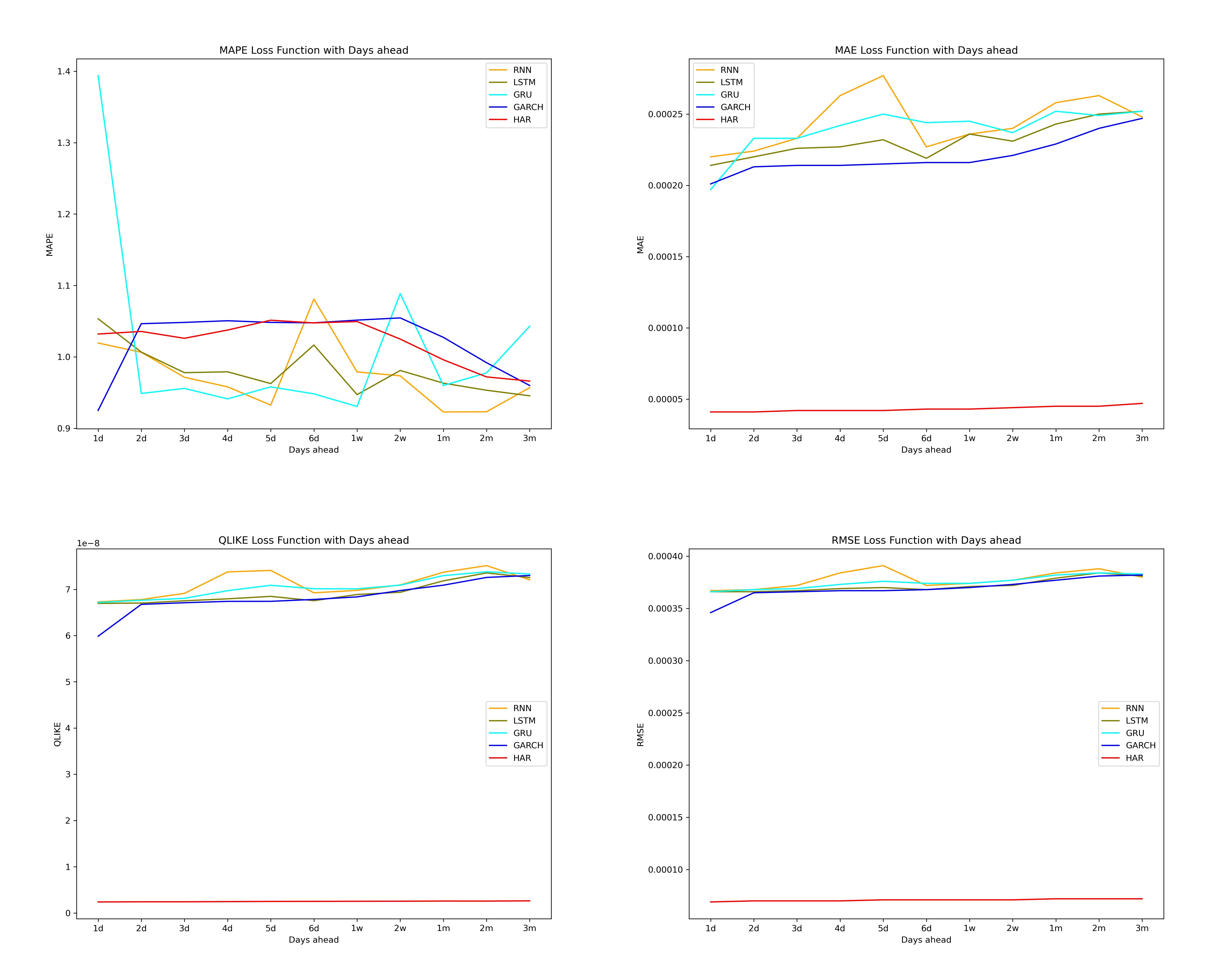}
\end{figure}

When using the RMSE loss function, the GARCH model demonstrates a significant advantage over recurrent neural network models when forecasting volatility for a single day ahead. However, as the forecast horizon extends and the number of days ahead increases, the GARCH model's advantage diminishes, displaying a gradual incremental increase in error. In contrast, the MAPE loss function shows a more erratic pattern, with no clear trend emerging over time, making it difficult to draw definitive conclusions regarding model performance based on this metric alone. 

Due to the inherent "black box" nature of machine learning models, recurrent neural networks (RNN), long short-term memory networks (LSTM), and gated recurrent units (GRU) may outperform the GARCH model in forward volatility forecasts. This is especially true when the forecast period is extended beyond a few days. The QLIKE loss function, known for its robustness, provides a clearer picture of the diminishing advantage of the GARCH model as the forecast horizon increases. Over time, the deep learning models begin to outperform GARCH based on the QLIKE metric, highlighting their potential in capturing complex volatility patterns in forward predictions.

Interestingly, throughout these comparisons, the HAR model consistently demonstrates a superior ability to forecast volatility, regardless of the number of days ahead. Its robust performance across multiple horizons suggests that it remains a highly effective tool for volatility forecasting, outperforming both GARCH and deep learning models in many scenarios. This stability in forecasting makes the HAR model a valuable benchmark for both short-term and long-term volatility predictions.
\begin{table}[]
    \footnotesize
    \centering
    \caption{Daily Rolling Window Forecasting Loss Functions Trend}
    \resizebox{\textwidth}{!}{%
    \begin{tabular}{llllllllllll}
    \textbf{Days ahead} & \textbf{1d} & \textbf{2d} & \textbf{3d} & \textbf{4d} & \textbf{5d} & \textbf{6d} & \textbf{1w} & \textbf{2w} & \textbf{1m} & \textbf{2m} & \textbf{3m} \\ \hline
    \multicolumn{12}{c}{\textbf{RMSE}}                                                                                                                                            \\ \hline
    \textbf{RNN}        & 0.0004      & 0.0004      & 0.0004      & 0.0004      & 0.0004      & 0.0004      & 0.0004      & 0.0004      & 0.0004      & 0.0004      & 0.0004      \\
    \textbf{LSTM}       & 0.0004      & 0.0004      & 0.0004      & 0.0004      & 0.0004      & 0.0004      & 0.0004      & 0.0004      & 0.0004      & 0.0004      & 0.0004      \\
    \textbf{GRU}        & 0.0004      & 0.0004      & 0.0004      & 0.0004      & 0.0004      & 0.0004      & 0.0004      & 0.0004      & 0.0004      & 0.0004      & 0.0004      \\
    \textbf{GARCH}      & 0.0003      & 0.0004      & 0.0004      & 0.0004      & 0.0004      & 0.0004      & 0.0004      & 0.0004      & 0.0004      & 0.0004      & 0.0004      \\
    \textbf{HAR}        & 0.0001      & 0.0001      & 0.0001      & 0.0001      & 0.0001      & 0.0001      & 0.0001      & 0.0001      & 0.0001      & 0.0001      & 0.0001      \\
    \multicolumn{12}{c}{\textbf{MAPE}}                                                                                                                                            \\ \hline
    \textbf{RNN}        & 1.0194      & 1.0065      & 0.9715      & 0.9581      & 0.9325      & 1.0810      & 0.9790      & 0.9735      & 0.9229      & 0.9231      & 0.9569      \\
    \textbf{LSTM}       & 1.0534      & 1.0068      & 0.9779      & 0.9791      & 0.9627      & 1.0166      & 0.9472      & 0.9810      & 0.9633      & 0.9533      & 0.9455      \\
    \textbf{GRU}        & 1.3938      & 0.9488      & 0.9559      & 0.9413      & 0.9580      & 0.9483      & 0.9306      & 1.0886      & 0.9600      & 0.9776      & 1.0428      \\
    \textbf{GARCH}      & 0.9250      & 1.0465      & 1.0483      & 1.0506      & 1.0483      & 1.0477      & 1.0515      & 1.0546      & 1.0274      & 0.9919      & 0.9600      \\
    \textbf{HAR}        & 1.0320      & 1.0357      & 1.0260      & 1.0376      & 1.0513      & 1.0476      & 1.0495      & 1.0250      & 0.9960      & 0.9720      & 0.9661      \\
    \multicolumn{12}{c}{\textbf{MAE}}                                                                                                                                             \\ \hline
    \textbf{RNN}        & 0.0002      & 0.0002      & 0.0002      & 0.0003      & 0.0003      & 0.0002      & 0.0002      & 0.0002      & 0.0003      & 0.0003      & 0.0002      \\
    \textbf{LSTM}       & 0.0002      & 0.0002      & 0.0002      & 0.0002      & 0.0002      & 0.0002      & 0.0002      & 0.0002      & 0.0002      & 0.0003      & 0.0003      \\
    \textbf{GRU}        & 0.0002      & 0.0002      & 0.0002      & 0.0002      & 0.0003      & 0.0002      & 0.0002      & 0.0002      & 0.0003      & 0.0002      & 0.0003      \\
    \textbf{GARCH}      & 0.0002      & 0.0002      & 0.0002      & 0.0002      & 0.0002      & 0.0002      & 0.0002      & 0.0002      & 0.0002      & 0.0002      & 0.0002      \\
    \textbf{HAR}        & 0.0000      & 0.0000      & 0.0000      & 0.0000      & 0.0000      & 0.0000      & 0.0000      & 0.0000      & 0.0000      & 0.0000      & 0.0000      \\
    \multicolumn{12}{c}{\textbf{QLIKE}}                                                                                                                                           \\ \hline
    \textbf{RNN}        & 6.73E-08    & 6.78E-08    & 6.91E-08    & 7.38E-08    & 7.41E-08    & 6.92E-08    & 6.98E-08    & 7.09E-08    & 7.37E-08    & 7.51E-08    & 7.21E-08    \\
    \textbf{LSTM}       & 6.70E-08    & 6.70E-08    & 6.75E-08    & 6.79E-08    & 6.85E-08    & 6.75E-08    & 6.89E-08    & 6.94E-08    & 7.18E-08    & 7.35E-08    & 7.25E-08    \\
    \textbf{GRU}        & 6.72E-08    & 6.76E-08    & 6.80E-08    & 6.97E-08    & 7.09E-08    & 7.01E-08    & 7.01E-08    & 7.09E-08    & 7.30E-08    & 7.38E-08    & 7.33E-08    \\
    \textbf{GARCH}      & 5.99E-08    & 6.68E-08    & 6.71E-08    & 6.74E-08    & 6.74E-08    & 6.78E-08    & 6.84E-08    & 6.97E-08    & 7.09E-08    & 7.26E-08    & 7.30E-08    \\
    \textbf{HAR}        & 2.39E-09    & 2.42E-09    & 2.42E-09    & 2.46E-09    & 2.50E-09    & 2.51E-09    & 2.53E-09    & 2.55E-09    & 2.58E-09    & 2.57E-09    & 2.63E-09    \\ \hline
    \end{tabular}%
    }

    \end{table}

\section{Conclusion}
In this paper, the econometric volatility models, particularly the HAR model, generally outperform recurrent neural network models in forecasting daily realized volatility for COMEX copper futures using a rolling window approach. Notably, the HAR model achieves a significant advantage with a QLIKE loss function value of 2.39E-09, which is an order of magnitude smaller than its competitors. The forecast results from the econometric models are notably smoother, as seen in the visualizations, while the deep learning models, such as RNN, LSTM, and GRU, demonstrate greater sensitivity to data changes, especially in periods of high market volatility. This higher sensitivity may reflect the deep learning models' ability to capture short-term fluctuations more rapidly than the econometric approaches.

In contrast, for higher-frequency rolling window forecasts of hourly realized volatility, the performance of deep learning models improves significantly. In this scenario, the recurrent neural network models achieve a QLIKE loss function value comparable to that of the HAR model, both reaching 3.43E-11. However, the GARCH model lags considerably, with a QLIKE loss function value of 5.19E-11. Despite this, the GARCH model’s prediction of extreme volatility events remains closer to the true values, as illustrated in the figures. This suggests that in high-frequency trading environments, investors should pay particular attention to the GARCH model’s extreme predictions, which could serve as an early warning signal for significant market movements. The deep learning models, on the other hand, show quicker adjustments to volatility changes, making them more suitable for real-time high-frequency forecasting.

As the forecast horizon for daily realized volatility extends, the predictive performance of deep learning models for COMEX copper futures gradually approaches that of the GARCH model. The increase in the QLIKE loss function for the recurrent neural network models is less pronounced compared to GARCH, indicating a more stable error rate as the forecast window extends. Furthermore, due to the inherent "black-box" nature of deep learning models, their loss function values in some experimental cases even fall below those of the GARCH model. However, despite these improvements, the HAR model retains a clear and significant advantage in forecasting daily volatility. Its consistent performance across different forecast horizons highlights its reliability, making it a superior model for daily volatility predictions in this context.

\clearpage
\nocite{*}
\bibliographystyle{apacite}
\bibliography{myref}

\begin{thebibliography}{}

\bibitem [\protect \citeauthoryear {%
Bucci%
}{%
Bucci%
}{%
{\protect \APACyear {2020}}%
}]{%
bucci2020realized}
\APACinsertmetastar {%
bucci2020realized}%
\begin{APACrefauthors}%
Bucci, A.%
\end{APACrefauthors}%
\unskip\
\newblock
\APACrefYearMonthDay{2020}{}{}.
\newblock
{\BBOQ}\APACrefatitle {Realized volatility forecasting with neural networks}
  {Realized volatility forecasting with neural networks}.{\BBCQ}
\newblock
\APACjournalVolNumPages{Journal of Financial Econometrics}{18}{3}{502--531}.
\PrintBackRefs{\CurrentBib}

\bibitem [\protect \citeauthoryear {%
Buncic%
\ \BBA {} Moretto%
}{%
Buncic%
\ \BBA {} Moretto%
}{%
{\protect \APACyear {2015}}%
}]{%
buncic2015forecasting}
\APACinsertmetastar {%
buncic2015forecasting}%
\begin{APACrefauthors}%
Buncic, D.%
\BCBT {}\ \BBA {} Moretto, C.%
\end{APACrefauthors}%
\unskip\
\newblock
\APACrefYearMonthDay{2015}{}{}.
\newblock
{\BBOQ}\APACrefatitle {Forecasting copper prices with dynamic averaging and
  selection models} {Forecasting copper prices with dynamic averaging and
  selection models}.{\BBCQ}
\newblock
\APACjournalVolNumPages{The North American Journal of Economics and
  Finance}{33}{}{1--38}.
\PrintBackRefs{\CurrentBib}

\bibitem [\protect \citeauthoryear {%
Colacito%
, Engle%
\BCBL {}\ \BBA {} Ghysels%
}{%
Colacito%
\ \protect \BOthers {.}}{%
{\protect \APACyear {2011}}%
}]{%
colacito2011component}
\APACinsertmetastar {%
colacito2011component}%
\begin{APACrefauthors}%
Colacito, R.%
, Engle, R\BPBI F.%
\BCBL {}\ \BBA {} Ghysels, E.%
\end{APACrefauthors}%
\unskip\
\newblock
\APACrefYearMonthDay{2011}{}{}.
\newblock
{\BBOQ}\APACrefatitle {A component model for dynamic correlations} {A component
  model for dynamic correlations}.{\BBCQ}
\newblock
\APACjournalVolNumPages{Journal of Econometrics}{164}{1}{45--59}.
\PrintBackRefs{\CurrentBib}

\bibitem [\protect \citeauthoryear {%
Conrad%
\ \BBA {} Kleen%
}{%
Conrad%
\ \BBA {} Kleen%
}{%
{\protect \APACyear {2020}}%
}]{%
conrad2020two}
\APACinsertmetastar {%
conrad2020two}%
\begin{APACrefauthors}%
Conrad, C.%
\BCBT {}\ \BBA {} Kleen, O.%
\end{APACrefauthors}%
\unskip\
\newblock
\APACrefYearMonthDay{2020}{}{}.
\newblock
{\BBOQ}\APACrefatitle {Two are better than one: Volatility forecasting using
  multiplicative component GARCH-MIDAS models} {Two are better than one:
  Volatility forecasting using multiplicative component garch-midas
  models}.{\BBCQ}
\newblock
\APACjournalVolNumPages{Journal of Applied Econometrics}{35}{1}{19--45}.
\PrintBackRefs{\CurrentBib}

\bibitem [\protect \citeauthoryear {%
D{\'\i}az%
, Hansen%
\BCBL {}\ \BBA {} Cabrera%
}{%
D{\'\i}az%
\ \protect \BOthers {.}}{%
{\protect \APACyear {2021}}%
}]{%
diaz2021economic}
\APACinsertmetastar {%
diaz2021economic}%
\begin{APACrefauthors}%
D{\'\i}az, J\BPBI D.%
, Hansen, E.%
\BCBL {}\ \BBA {} Cabrera, G.%
\end{APACrefauthors}%
\unskip\
\newblock
\APACrefYearMonthDay{2021}{}{}.
\newblock
{\BBOQ}\APACrefatitle {Economic drivers of commodity volatility: The case of
  copper} {Economic drivers of commodity volatility: The case of
  copper}.{\BBCQ}
\newblock
\APACjournalVolNumPages{Resources Policy}{73}{}{}.
\PrintBackRefs{\CurrentBib}

\bibitem [\protect \citeauthoryear {%
Duvhammar%
}{%
Duvhammar%
}{%
{\protect \APACyear {2018}}%
}]{%
duvhammar2018volatility}
\APACinsertmetastar {%
duvhammar2018volatility}%
\begin{APACrefauthors}%
Duvhammar, M.%
\end{APACrefauthors}%
\unskip\
\newblock
\APACrefYearMonthDay{2018}{}{}.
\newblock
{\BBOQ}\APACrefatitle {Volatility of copper prices and the effect of real
  interest rate changes} {Volatility of copper prices and the effect of real
  interest rate changes}.{\BBCQ}
\newblock
\APACjournalVolNumPages{Swedish University of Agricultural Sciences}{}{}{}.
\PrintBackRefs{\CurrentBib}

\bibitem [\protect \citeauthoryear {%
Elder%
, Miao%
\BCBL {}\ \BBA {} Ramchander%
}{%
Elder%
\ \protect \BOthers {.}}{%
{\protect \APACyear {2012}}%
}]{%
elder2012impact}
\APACinsertmetastar {%
elder2012impact}%
\begin{APACrefauthors}%
Elder, J.%
, Miao, H.%
\BCBL {}\ \BBA {} Ramchander, S.%
\end{APACrefauthors}%
\unskip\
\newblock
\APACrefYearMonthDay{2012}{}{}.
\newblock
{\BBOQ}\APACrefatitle {Impact of macroeconomic news on metal futures} {Impact
  of macroeconomic news on metal futures}.{\BBCQ}
\newblock
\APACjournalVolNumPages{Journal of Banking \& Finance}{36}{1}{51--65}.
\PrintBackRefs{\CurrentBib}

\bibitem [\protect \citeauthoryear {%
Engle%
, Ghysels%
\BCBL {}\ \BBA {} Sohn%
}{%
Engle%
\ \protect \BOthers {.}}{%
{\protect \APACyear {2008}}%
}]{%
engle2008economic}
\APACinsertmetastar {%
engle2008economic}%
\begin{APACrefauthors}%
Engle, R\BPBI F.%
, Ghysels, E.%
\BCBL {}\ \BBA {} Sohn, B.%
\end{APACrefauthors}%
\unskip\
\newblock
\APACrefYearMonthDay{2008}{}{}.
\newblock
{\BBOQ}\APACrefatitle {On the economic sources of stock market volatility} {On
  the economic sources of stock market volatility}.{\BBCQ}
\newblock
\BIn{} \APACrefbtitle {AFA 2008 New Orleans Meetings Paper.} {Afa 2008 new
  orleans meetings paper.}
\PrintBackRefs{\CurrentBib}

\bibitem [\protect \citeauthoryear {%
Engle%
, Ghysels%
\BCBL {}\ \BBA {} Sohn%
}{%
Engle%
\ \protect \BOthers {.}}{%
{\protect \APACyear {2013}}%
}]{%
engle2013stock}
\APACinsertmetastar {%
engle2013stock}%
\begin{APACrefauthors}%
Engle, R\BPBI F.%
, Ghysels, E.%
\BCBL {}\ \BBA {} Sohn, B.%
\end{APACrefauthors}%
\unskip\
\newblock
\APACrefYearMonthDay{2013}{}{}.
\newblock
{\BBOQ}\APACrefatitle {Stock market volatility and macroeconomic fundamentals}
  {Stock market volatility and macroeconomic fundamentals}.{\BBCQ}
\newblock
\APACjournalVolNumPages{Review of Economics and Statistics}{95}{3}{776--797}.
\PrintBackRefs{\CurrentBib}

\bibitem [\protect \citeauthoryear {%
Ghysels%
, Santa-Clara%
\BCBL {}\ \BBA {} Valkanov%
}{%
Ghysels%
\ \protect \BOthers {.}}{%
{\protect \APACyear {2004}}%
}]{%
ghysels2004midas}
\APACinsertmetastar {%
ghysels2004midas}%
\begin{APACrefauthors}%
Ghysels, E.%
, Santa-Clara, P.%
\BCBL {}\ \BBA {} Valkanov, R.%
\end{APACrefauthors}%
\unskip\
\newblock
\APACrefYearMonthDay{2004}{}{}.
\newblock
{\BBOQ}\APACrefatitle {The MIDAS touch: Mixed data sampling regression models}
  {The midas touch: Mixed data sampling regression models}.{\BBCQ}
\newblock

\PrintBackRefs{\CurrentBib}

\bibitem [\protect \citeauthoryear {%
Guzm{\'a}n~Barros%
\ \BBA {} Silva%
}{%
Guzm{\'a}n~Barros%
\ \BBA {} Silva%
}{%
{\protect \APACyear {2018}}%
}]{%
guzman2018copper}
\APACinsertmetastar {%
guzman2018copper}%
\begin{APACrefauthors}%
Guzm{\'a}n~Barros, J\BPBI I.%
\BCBT {}\ \BBA {} Silva, E.%
\end{APACrefauthors}%
\unskip\
\newblock
\APACrefYearMonthDay{2018}{}{}.
\newblock
{\BBOQ}\APACrefatitle {Copper price determination: fundamentals versus
  non-fundamentals} {Copper price determination: fundamentals versus
  non-fundamentals}.{\BBCQ}
\newblock

\PrintBackRefs{\CurrentBib}

\bibitem [\protect \citeauthoryear {%
Hammoudeh%
\ \BBA {} Yuan%
}{%
Hammoudeh%
\ \BBA {} Yuan%
}{%
{\protect \APACyear {2008}}%
}]{%
hammoudeh2008metal}
\APACinsertmetastar {%
hammoudeh2008metal}%
\begin{APACrefauthors}%
Hammoudeh, S.%
\BCBT {}\ \BBA {} Yuan, Y.%
\end{APACrefauthors}%
\unskip\
\newblock
\APACrefYearMonthDay{2008}{}{}.
\newblock
{\BBOQ}\APACrefatitle {Metal volatility in presence of oil and interest rate
  shocks} {Metal volatility in presence of oil and interest rate
  shocks}.{\BBCQ}
\newblock
\APACjournalVolNumPages{Energy Economics}{30}{2}{606--620}.
\PrintBackRefs{\CurrentBib}

\bibitem [\protect \citeauthoryear {%
Liu%
, Yang%
\BCBL {}\ \BBA {} Ruan%
}{%
Liu%
\ \protect \BOthers {.}}{%
{\protect \APACyear {2019}}%
}]{%
liu2019impact}
\APACinsertmetastar {%
liu2019impact}%
\begin{APACrefauthors}%
Liu, R.%
, Yang, J.%
\BCBL {}\ \BBA {} Ruan, C\BPBI Y.%
\end{APACrefauthors}%
\unskip\
\newblock
\APACrefYearMonthDay{2019}{}{}.
\newblock
{\BBOQ}\APACrefatitle {The impact of macroeconomic news on Chinese futures}
  {The impact of macroeconomic news on chinese futures}.{\BBCQ}
\newblock
\APACjournalVolNumPages{International Journal of Financial Studies}{7}{4}{63}.
\PrintBackRefs{\CurrentBib}

\bibitem [\protect \citeauthoryear {%
Patton%
}{%
Patton%
}{%
{\protect \APACyear {2011}}%
}]{%
patton2011volatility}
\APACinsertmetastar {%
patton2011volatility}%
\begin{APACrefauthors}%
Patton, A\BPBI J.%
\end{APACrefauthors}%
\unskip\
\newblock
\APACrefYearMonthDay{2011}{}{}.
\newblock
{\BBOQ}\APACrefatitle {Volatility forecast comparison using imperfect
  volatility proxies} {Volatility forecast comparison using imperfect
  volatility proxies}.{\BBCQ}
\newblock
\APACjournalVolNumPages{Journal of Econometrics}{160}{1}{246--256}.
\PrintBackRefs{\CurrentBib}

\bibitem [\protect \citeauthoryear {%
Rahimikia%
\ \BBA {} Poon%
}{%
Rahimikia%
\ \BBA {} Poon%
}{%
{\protect \APACyear {2020}}%
}]{%
rahimikia2020machine}
\APACinsertmetastar {%
rahimikia2020machine}%
\begin{APACrefauthors}%
Rahimikia, E.%
\BCBT {}\ \BBA {} Poon, S\BHBI H.%
\end{APACrefauthors}%
\unskip\
\newblock
\APACrefYearMonthDay{2020}{}{}.
\newblock
{\BBOQ}\APACrefatitle {Machine learning for realised volatility forecasting}
  {Machine learning for realised volatility forecasting}.{\BBCQ}
\newblock
\APACjournalVolNumPages{Available at SSRN}{}{}{}.
\PrintBackRefs{\CurrentBib}

\bibitem [\protect \citeauthoryear {%
Robert%
\ \protect \BOthers {.}}{%
Robert%
\ \protect \BOthers {.}}{%
{\protect \APACyear {2002}}%
}]{%
robert2002dynamic}
\APACinsertmetastar {%
robert2002dynamic}%
\begin{APACrefauthors}%
Robert, E\BPBI F.%
\BCBT {}\ \BOthersPeriod {.}
\end{APACrefauthors}%
\unskip\
\newblock
\APACrefYearMonthDay{2002}{}{}.
\newblock
{\BBOQ}\APACrefatitle {Dynamic conditional correlation: A simple class of
  multivariate generalized autoregressive conditional hetroscedasticity models}
  {Dynamic conditional correlation: A simple class of multivariate generalized
  autoregressive conditional hetroscedasticity models}.{\BBCQ}
\newblock
\APACjournalVolNumPages{Journal of Business and Economic
  Statistics}{20}{3}{339--350}.
\PrintBackRefs{\CurrentBib}

\bibitem [\protect \citeauthoryear {%
Rodikov%
\ \BBA {} Antulov-Fantulin%
}{%
Rodikov%
\ \BBA {} Antulov-Fantulin%
}{%
{\protect \APACyear {2022}}%
}]{%
rodikov2022can}
\APACinsertmetastar {%
rodikov2022can}%
\begin{APACrefauthors}%
Rodikov, G.%
\BCBT {}\ \BBA {} Antulov-Fantulin, N.%
\end{APACrefauthors}%
\unskip\
\newblock
\APACrefYearMonthDay{2022}{}{}.
\newblock
{\BBOQ}\APACrefatitle {Can LSTM outperform volatility-econometric models?} {Can
  lstm outperform volatility-econometric models?}{\BBCQ}
\newblock
\APACjournalVolNumPages{arXiv preprint arXiv:2202.11581}{}{}{}.
\PrintBackRefs{\CurrentBib}

\bibitem [\protect \citeauthoryear {%
Sadorsky%
}{%
Sadorsky%
}{%
{\protect \APACyear {2014}}%
}]{%
sadorsky2014modeling}
\APACinsertmetastar {%
sadorsky2014modeling}%
\begin{APACrefauthors}%
Sadorsky, P.%
\end{APACrefauthors}%
\unskip\
\newblock
\APACrefYearMonthDay{2014}{}{}.
\newblock
{\BBOQ}\APACrefatitle {Modeling volatility and correlations between emerging
  market stock prices and the prices of copper, oil and wheat} {Modeling
  volatility and correlations between emerging market stock prices and the
  prices of copper, oil and wheat}.{\BBCQ}
\newblock
\APACjournalVolNumPages{Energy Economics}{43}{}{72--81}.
\PrintBackRefs{\CurrentBib}

\bibitem [\protect \citeauthoryear {%
Smith%
\ \BBA {} Bracker%
}{%
Smith%
\ \BBA {} Bracker%
}{%
{\protect \APACyear {2003}}%
}]{%
smith2003forecasting}
\APACinsertmetastar {%
smith2003forecasting}%
\begin{APACrefauthors}%
Smith, K\BPBI L.%
\BCBT {}\ \BBA {} Bracker, K.%
\end{APACrefauthors}%
\unskip\
\newblock
\APACrefYearMonthDay{2003}{}{}.
\newblock
{\BBOQ}\APACrefatitle {Forecasting changes in copper futures volatility with
  GARCH models using an iterated algorithm} {Forecasting changes in copper
  futures volatility with garch models using an iterated algorithm}.{\BBCQ}
\newblock
\APACjournalVolNumPages{Review of Quantitative Finance and
  Accounting}{20}{3}{245--265}.
\PrintBackRefs{\CurrentBib}

\bibitem [\protect \citeauthoryear {%
Sreenu%
\ \BBA {} Rao%
}{%
Sreenu%
\ \BBA {} Rao%
}{%
{\protect \APACyear {2021}}%
}]{%
sreenu2021macroeconomic}
\APACinsertmetastar {%
sreenu2021macroeconomic}%
\begin{APACrefauthors}%
Sreenu, N.%
\BCBT {}\ \BBA {} Rao, K.%
\end{APACrefauthors}%
\unskip\
\newblock
\APACrefYearMonthDay{2021}{}{}.
\newblock
{\BBOQ}\APACrefatitle {The macroeconomic variables impact on commodity futures
  volatility: A study on Indian markets} {The macroeconomic variables impact on
  commodity futures volatility: A study on indian markets}.{\BBCQ}
\newblock
\APACjournalVolNumPages{Cogent Business \& Management}{8}{1}{}.
\PrintBackRefs{\CurrentBib}

\bibitem [\protect \citeauthoryear {%
Wang%
, Wang%
\BCBL {}\ \BBA {} Wang%
}{%
Wang%
\ \protect \BOthers {.}}{%
{\protect \APACyear {2020}}%
}]{%
wang2020research}
\APACinsertmetastar {%
wang2020research}%
\begin{APACrefauthors}%
Wang, L.%
, Wang, H.%
\BCBL {}\ \BBA {} Wang, J.%
\end{APACrefauthors}%
\unskip\
\newblock
\APACrefYearMonthDay{2020}{}{}.
\newblock
{\BBOQ}\APACrefatitle {Research on the Influence of Economic Policy Uncertainty
  on the Supply Chain Finance} {Research on the influence of economic policy
  uncertainty on the supply chain finance}.{\BBCQ}
\newblock
\BIn{} \APACrefbtitle {E3S Web of Conferences} {E3s web of conferences}\
  (\BVOL~214).
\PrintBackRefs{\CurrentBib}

\bibitem [\protect \citeauthoryear {%
Xiao%
, Su%
\BCBL {}\ \BBA {} Ayub%
}{%
Xiao%
\ \protect \BOthers {.}}{%
{\protect \APACyear {2022}}%
}]{%
xiao2022economic}
\APACinsertmetastar {%
xiao2022economic}%
\begin{APACrefauthors}%
Xiao, D.%
, Su, J.%
\BCBL {}\ \BBA {} Ayub, B.%
\end{APACrefauthors}%
\unskip\
\newblock
\APACrefYearMonthDay{2022}{}{}.
\newblock
{\BBOQ}\APACrefatitle {Economic policy uncertainty and commodity market
  volatility: implications for economic recovery} {Economic policy uncertainty
  and commodity market volatility: implications for economic recovery}.{\BBCQ}
\newblock
\APACjournalVolNumPages{Environmental Science and Pollution
  Research}{}{}{1--12}.
\PrintBackRefs{\CurrentBib}

\bibitem [\protect \citeauthoryear {%
Xiong%
, Nichols%
\BCBL {}\ \BBA {} Shen%
}{%
Xiong%
\ \protect \BOthers {.}}{%
{\protect \APACyear {2015}}%
}]{%
xiong2015deep}
\APACinsertmetastar {%
xiong2015deep}%
\begin{APACrefauthors}%
Xiong, R.%
, Nichols, E\BPBI P.%
\BCBL {}\ \BBA {} Shen, Y.%
\end{APACrefauthors}%
\unskip\
\newblock
\APACrefYearMonthDay{2015}{}{}.
\newblock
{\BBOQ}\APACrefatitle {Deep learning stock volatility with google domestic
  trends} {Deep learning stock volatility with google domestic trends}.{\BBCQ}
\newblock
\APACjournalVolNumPages{arXiv preprint arXiv:1512.04916}{}{}{}.
\PrintBackRefs{\CurrentBib}

\bibitem [\protect \citeauthoryear {%
C.~Zhang%
, Zhang%
, Cucuringu%
\BCBL {}\ \BBA {} Qian%
}{%
C.~Zhang%
\ \protect \BOthers {.}}{%
{\protect \APACyear {2022}}%
}]{%
zhang2022volatility}
\APACinsertmetastar {%
zhang2022volatility}%
\begin{APACrefauthors}%
Zhang, C.%
, Zhang, Y.%
, Cucuringu, M.%
\BCBL {}\ \BBA {} Qian, Z.%
\end{APACrefauthors}%
\unskip\
\newblock
\APACrefYearMonthDay{2022}{}{}.
\newblock
{\BBOQ}\APACrefatitle {Volatility forecasting with machine learning and
  intraday commonality} {Volatility forecasting with machine learning and
  intraday commonality}.{\BBCQ}
\newblock
\APACjournalVolNumPages{arXiv preprint arXiv:2202.08962}{}{}{}.
\PrintBackRefs{\CurrentBib}

\bibitem [\protect \citeauthoryear {%
Y.~Zhang%
\ \BBA {} Wang%
}{%
Y.~Zhang%
\ \BBA {} Wang%
}{%
{\protect \APACyear {2022}}%
}]{%
zhang2022covid}
\APACinsertmetastar {%
zhang2022covid}%
\begin{APACrefauthors}%
Zhang, Y.%
\BCBT {}\ \BBA {} Wang, R.%
\end{APACrefauthors}%
\unskip\
\newblock
\APACrefYearMonthDay{2022}{}{}.
\newblock
{\BBOQ}\APACrefatitle {COVID-19 impact on commodity futures volatilities}
  {Covid-19 impact on commodity futures volatilities}.{\BBCQ}
\newblock
\APACjournalVolNumPages{Finance Research Letters}{47}{}{}.
\PrintBackRefs{\CurrentBib}

\bibitem [\protect \citeauthoryear {%
Zheng%
, Zhou%
\BCBL {}\ \BBA {} Wen%
}{%
Zheng%
\ \protect \BOthers {.}}{%
{\protect \APACyear {2021}}%
}]{%
zheng2021asymmetric}
\APACinsertmetastar {%
zheng2021asymmetric}%
\begin{APACrefauthors}%
Zheng, Y.%
, Zhou, M.%
\BCBL {}\ \BBA {} Wen, F.%
\end{APACrefauthors}%
\unskip\
\newblock
\APACrefYearMonthDay{2021}{}{}.
\newblock
{\BBOQ}\APACrefatitle {Asymmetric effects of oil shocks on carbon allowance
  price: evidence from China} {Asymmetric effects of oil shocks on carbon
  allowance price: evidence from china}.{\BBCQ}
\newblock
\APACjournalVolNumPages{Energy Economics}{97}{}{}.
\PrintBackRefs{\CurrentBib}

\end{thebibliography}
\end{document}